\documentclass[12pt]{iopart}

\usepackage{graphicx}
\usepackage{dcolumn}
\usepackage{bm}
\usepackage{bbm}
\usepackage{epsfig}
\usepackage{mathrsfs}
\usepackage{stmaryrd}
\usepackage{color}
\usepackage{dsfont}
\usepackage{iopams}
\usepackage{psfrag}
\usepackage[utf8]{inputenc}

\newcommand{\bra}[1]{\langle\,{#1}\, |}
\newcommand{\ket}[1]{|\,{#1}\,\rangle}

\newcommand{\aver}[1]{\langle\!\langle\, {#1}\,\rangle\!\rangle}

\newcommand{\Imag}{\mbox{Im}}
\newcommand{\Real}{\mbox{Re}}

\newcommand{\dt}{\mbox{d}t}
\newcommand{\dW}{\mbox{d}W}
\newcommand{\dd}{\mbox{d}}
\newcommand{\opH}{\boldsymbol{H}}

\newcommand{\cref}[1]{chapter~\ref{#1}}
\newcommand{\Cref}[1]{Chapter~\ref{#1}}

\newcommand{\rA}{R}
\newcommand{\rB}{S}
\newcommand{\rC}{T}
\newcommand{\rAqm}{R^{\rm qm}}
\newcommand{\rBqm}{S^{\rm qm}}
\newcommand{\rCqm}{T^{\rm qm}}
\newcommand{\dotrAqm}{\dot{R}^{\rm qm}}
\newcommand{\dotrBqm}{\dot{S}^{\rm qm}}
\newcommand{\dotrCqm}{\dot{T}^{\rm qm}}

\begin{document}

\title[Classical master equation for excitonic transport]{A classical master equation for excitonic transport under the influence of an environment
}
\author{Alexander Eisfeld$^{1,2}$ and John~S.~Briggs$^{1}$ }
\address{$^{1}$Max Planck Institute for the Physics of Complex Systems,
N\"othnitzer Strasse 38, 01187 Dresden, Germany\\
$^{2}$Department of Chemistry and Chemical Biology
Harvard University
12 Oxford Street,
Cambridge, MA 02138}
\ead{eisfeld@mpipks-dresden.mpg.de}

\begin{abstract}
 In a previous paper [Phys.~Rev.~E 83, 051911] we have shown that the results of a quantum-mechanical calculation of electronic energy transfer (EET) over aggregates of coupled monomers can 
 be described also by a model of interacting classical electric dipoles in a weak-coupling approximation, which we referred to as the realistic coupling approximation (RCA). The method was illustrated by EET on a simple linear chain of molecules and also by energy transfer on the Fenna-Matthews-Olson (FMO) complex relevant for photosynthesis. The study was limited to electronic degrees of freedom since this is the origin of coherent EET in the quantum case. Nevertheless, more realistic models of EET require the inclusion of the de-cohering effects of coupling to an environment, when the molecular aggregate becomes an open quantum system. Here we consider the quantum description of EET on a linear chain and on the FMO complex, incorporating environment coupling and construct the classical version of the same systems in the density matrix formalism. The close agreement  of the exact quantum and exact classical results in the RCA is demonstrated and justified  analytically. This lends further support to the conclusion that the coherence properties of EET in the FMO complex is evident at the classical level and should not be ascribed as solely due to quantum effects.
 
\end{abstract}
\maketitle

\section{Introduction}
In a previous communication \cite{BrEi11_051911_} (to be denoted as paper 1) we studied theoretically the process of electronic energy transfer (EET) on molecular aggregates. The aim was to demonstrate that the coherent transfer arising from  an entangled aggregate wavefunction (an exciton) in the quantum case, in the approximation that there is no exchange of electrons between monomers, is reproduced by a classical model of the aggregate as an assembly of electrical dipoles.  The quantum/classical equivalence is valid in what we called the "realistic coupling approximation" (RCA). This is a weak-coupling approximation in that the strength of the dipole-dipole interaction that effects the transfer is considered small compared to typical  electronic excitation energies, so that the monomers largely retain their character upon excitation of the aggregate. Practically this implies that the exciton bandwidth and the average spread of monomeric transition energies are both small compared to the mean electronic transition energy. Happily these criteria do pertain in many dye aggregates and also in the photosynthetic unit, so that the RCA is valid and one expects the classical model to give results in agreement with the quantum theory.

As specific example, we considered first the transfer of energy along a chain of identical monomers where, in the approximation that only nearest-neighbors interact, an analytical solution for the transfer probability is possible. This solution predicts oscillatory monomer excitation probability in time and a constant velocity EET along the chain from an initially-excited monomer. The constant velocity and the oscillating nature of the transfer probability are signatures of fully-coherent propagation. Significantly it was shown, by numerical solution of the full classical equations for the same coupling, that classical interacting dipoles lead to exactly the same coherent  transfer as in the quantum case. 
As a second model we considered EET on the Fenna-Matthews-Olson (FMO) photosynthetic complex, where the local transition energies of bacteriochlorophyll molecules on different sites are unequal. Here again, using realistic transition energies and coupling strengths, we were able to show the equivalence of classical and quantum dynamics in deciding the coherence of EET. 
The equivalence in this example is of particular significance since here the coherence of EET  has been attributed \cite{SaIsFl10_462_} as arising solely from the entanglement properties of the aggregate electronic wavefunction. Hence, were this to be true, one would not expect such coherence to be present in the results of a purely classical treatment.

In the above examples only electronic degrees of freedom were considered, since the emphasis was on coherent EET and the electronic excitation is the seat of such coherence. Nevertheless, particularly with reference to the FMO complex, in any real molecular aggregate the internal electronic degrees of freedom experience interaction with internal and external vibrational modes and electromagnetic interaction with the surrounding solvent. In absorption and emission of photons such interactions manifest themselves obviously in broadening and shifting of spectral bands. In EET there are more subtle manifestations in the de-phasing, de-cohering and sometimes dissipating effects on propagation of absorbed light energy. Hence, to obtain a more realistic overall picture and to further test the classical model of EET we feel it essential to include the interactions with the surroundings. This enlargement of the theoretical model, to consider the molecular aggregate as an open quantum or classical system, is the subject of this paper.

The development of the paper proceeds as follows. In section \ref{sec:QM} we consider the quantum case of a molecular aggregate in which the coupling to the environment is represented by interaction of electronic degrees of freedom with external stochastic fluctuations. These lead to de-phasing and time-varying electronic transition energies. Here we adopt the usual density matrix formulation leading  to a Lindblad-type master equation which is equivalent to that derived by Haken, Reineker and Strobl \cite{HaRe71_253_,HaSt73_135_} arising from Markovian environment fluctuations. This will be denoted as the HRS equation. 
This model (and its variants) has been used extensively to describe exciton transport in molecular crystals, molecular aggregates and photosynthetic complexes (see e.g.\ Refs.~\cite{LiVaGr97_7343_,RaKnKe79_197_,ReWaNe93_715_,SuEnRe06_337_,CaChDa09_105106_,ReMoKa09_033003_})

In section \ref{sec:class_mech} we apply the same physical assumptions to derive a \emph{classical} density matrix equation of a similar, but not exactly equivalent, form to the quantum equation. In an appendix, the equivalence of the classical  formulation in the RCA  to the quantum version is proved. The proof is most transparent using the stochastic Schr\"odinger quantum equation (which is equivalent to the HRS equation).

In section \ref{sec:RCA_test} the two cases already considered in paper 1 as 'bare' electronic systems are re-calculated including the effects of environment coupling. The main effect of this coupling is to damp out oscillations in occupation probabilities and to slow the rate of EET. Perhaps more importantly, by plotting density matrix elements we show that coherences between different sites are suppressed also.  In the case of the FMO complex, our model can now be considered a realistic representation of the main features of EET and yet, significantly, the exact purely classical model including environmental effects again gives results hardly distinguishable from the fully quantum results, including the coherences between different sites.

\section{Quantum Mechanics}
\label{sec:QM}

The excitonic part of the aggregate is described by the Hamiltonian
$
{\mathbf H_{\rm ex}}  ={\mathbf H_0} + {\mathbf V}
$
where $\mathbf H_0$ is the sum of the Hamiltonians of non-interacting monomers and $\mathbf V$ is the total potential energy of the pairwise interactions between monomers.
Since we consider the propagation of a single electronic excitation along the aggregate, we expand the Hamiltonian with respect to  states $\ket{\pi_n}$  in which monomer $n$ is electronically excited and all other monomers are in their ground state.
In this basis one has
\begin{equation}
\label{eq:H_ex}
\opH_{\rm ex}=\sum_{n}\epsilon_n \ket{\pi_n}\bra{\pi_n} +\sum_{n,m}V_{nm} \ket{\pi_n}\bra{\pi_m}  
\end{equation}
where  $\epsilon_n$ is the single-monomer transition energy and  the full aggregate ground-state energy is set to zero.
The matrix-element $V_{nm}$ describes excitation transfer between site $n$ and $m$.

In the following we are interested in the dynamics of the (reduced) density matrix $\rho(t)$ of the electronic system when the electronic excitation interacts with an environment.
We adopt a particular simple model where the dynamics of the density matrix $\rho(t)$ is determined by a Lindblad Master equation of the form 
\begin{equation}
  \label{eq:dot_rho_nm}
   \dot\rho_{nm}(t)=\mathcal{H}[\rho]_{nm}+\mathcal{L}[\rho]_{nm} .  
\end{equation}
where 
\begin{eqnarray}
\mathcal{H}[\rho]_{nm}&=&- \frac{i}{\hbar} [\opH_{\rm ex},\rho(t)]_{nm}\\
&=&-\frac{i}{\hbar}(\epsilon_n-\epsilon_m)\rho_{nm}
- \frac{i}{\hbar}\sum_{\ell}(V_{n\ell}\rho_{\ell m}- V_{\ell m}\rho_{n \ell})
 \label{eq:dot_rho_nm_detail}
\end{eqnarray}
and the last line follows from \eref{eq:H_ex}.
The interaction with the environment is contained in $\mathcal{L}[\rho]_{nm}$ 
which for simplicity we take to be
\begin{equation}
\label{eq:L_pure_deph}
\mathcal{L}[\rho]_{nm}=-\Big(\frac{1}{2}(\gamma_n+\gamma_m) -\sqrt{\gamma_n\gamma_m}\delta_{nm}\Big)\rho_{nm}
\end{equation}
i.e.\ we consider pure dephasing with dephasing rates $\gamma_n$.

The extension to the  general HRS master equation is straightforward. 
Also generalizations of the HRS model, as in Ref.~\cite{Ca85_101_,BlSi78_3589_,SzBa86_179_,WuKn98_359_}, can be treated similarly.
In the Conclusion we will discuss this point in more detail.

 In the following we will derive a classical equation which in RCA  is equivalent to Eq.~(\ref{eq:dot_rho_nm}). 
To this end we first note that the master equation (\ref{eq:dot_rho_nm}) is equivalent (following the treatment of HRS \cite{HaRe71_253_,HaSt73_135_})  to a stochastic Schr\"odinger equation $\partial_t\ket{\psi(t)}=-i\opH(t) \ket{\psi(t)}$ with local Markovian fluctuations of the site energies.
These fluctuations can be merged together with the transition energy of the monomers to obtain a stochastic Hamiltonian
\begin{equation}
\label{eq:Ham_deph_stoch}
\opH(t)=\sum_{n}\epsilon_n(t) \ket{\pi_n}\bra{\pi_n} +\sum_{n,m}V_{nm} \ket{\pi_n}\bra{\pi_m}  
\end{equation} 
where the fluctuations in the transition energies $\epsilon_n(t)$ have the properties of real Gaussian Markov processes  fulfilling
\begin{eqnarray}
\aver{\epsilon_{n}(t)}=\epsilon_n
\label{eq:qm_statistics_mean}\\
\aver{\epsilon_{n}(t)\epsilon_{m}(t')}= \hbar^2 \gamma_{n}\delta_{nm} \delta(t-t')
\label{eq:qm_statistics_var}
\end{eqnarray}
where $\aver{\cdots}$ denotes the averaging over many realizations of  the stochastic processes.
The master equation \eref{eq:dot_rho_nm} is then obtained by taking the time derivative of $\rho(t)=\aver{\ket{\psi(t)}\bra{\psi(t)}}$. 
We note that we have restricted ourselves to Gaussian Markov processes, since then one can derive a simple master equation. The arguments presented below on the validity of the RCA approximation are applicable to more general stochastic processes which do not have to be Gaussian (as long as the second moment exists) and can also be correlated in time (non-Markovian). 

The stochastic 'unravelling' \eref{eq:Ham_deph_stoch} will be used in the next section to make the connection to the classical case. In particular we will take the frequency of the classical oscillators to obey the same statistical properties as the quantum transition energies.
The resulting classical stochastic equation will then be used to construct a 'classical master equation'.

\section{Classical mechanics}
\label{sec:class_mech}
As in our previous publications \cite{BrEi11_051911_,BrEi11__a} in the classical case we consider that the coupled quantum transition dipoles are modelled by classical oscillators in the same geometry as the transition dipoles of the quantum mechanical model. 
The frequencies of the classical oscillators are associated with the quantum energies via  $ \omega_n=\epsilon_n/\hbar$. 
To arrive at a density matrix description corresponding to \eref{eq:dot_rho_nm} we make use of the stochastic representation introduced in the previous section.
Thus we assume that the frequency of the classical oscillators is a stochastic quantity with
\begin{equation}
\label{eq:omega(t)}
 \omega_n(t)=\epsilon_n(t) /\hbar
\end{equation}
and the same statistical properties  as given by \eref{eq:qm_statistics_mean} and \eref{eq:qm_statistics_var}.

The Hamilton equations of motion for linearly-interacting oscillators of {\it time-dependent} frequency $\omega_n(t)$ as defined above in \eref{eq:omega(t)}, are,
\begin{eqnarray}
\label{eq:classical_eq_motion_x}
\dot{x}_n=&\omega_n(t) p_n
\\
\label{eq:classical_eq_motion_p}
\dot{p}_n= &- \omega_n(t)x_n -\sum_m \tilde{K}_{nm} x_m 
\end{eqnarray}
where $x_n$ and $p_n$ are the dimensionless position and momentum of the $n$th oscillator respectively.
The $\tilde{K}_{nm}$ are coupling coefficients which are related to the quantum mechanical couplings by \cite{BrEi11_051911_,BrEi11__a}
\begin{equation}
\tilde{K}_{nm}=2 \frac{V_{nm}}{\hbar}.
\end{equation}

To connect to the quantum equations, we 
introduce the dimensionless complex amplitude 
\begin{equation}
{z}_n(t)={x}_n(t)+i {p}_n(t)
\end{equation}
and obtain the coupled equations
\begin{equation}
\label{eq:dot_z_final}
\dot{z}_n=-i \omega_n(t) z_n - i\sum_m \frac{V_{nm}}{\hbar} 2 \Real (z_m)
\end{equation}
These equations can be viewed as a set of coupled ``Kubo-oscillators'' \cite{Ku54_935_,Fo78_179_}.
Note that here $2 \Real (z_m)$ appears as coupling in the equation for the amplitudes. 
As shown in \cite{BrEi11_051911_} in an equation which would be fully equivalent to quantum mechanics this term would be replaced by $z_m$.

\subsection{Classical density operator}
\label{sec:class_dens_op}
To make contact with the pure dephasing master equation \eref{eq:dot_rho_nm}  we will take a closer look at the products
\begin{equation}
\label{eq:tilde(sigma)}
\tilde{\sigma}_{nm}(t)=z_{n}(t)z^{*}_m(t)
\end{equation}
which, as we will show below (in RCA) resemble the quantum mechanical density operator matrix elements.
In the following derivation some care has to be taken due to the stochastic nature of $\omega_n(t)$.
We interpret \eref{eq:dot_z_final} to be a stochastic Schr\"odinger equation in the Stratonowich form (see e.g.\ Refs.~\cite{Ka81_175_,Fo78_179_,HaEz80_41_}).
In the following derivation we use the Ito calculus (see e.g.\ Ref.~\cite{Ka81_175_}) and write \eref{eq:dot_z_final} in its Ito form 
\begin{eqnarray}
\label{eq:dz}
{\rm d} z_n=&\Big(-i \omega_nz_n- i \sum_m{\tilde{K}}_{nm}\Real{z_m} - \frac{\gamma_n}{2}z_n   \Big) \dt + \sqrt{\gamma_n}\, z_n\, \dW_n
\end{eqnarray}
Here ${\rm d} z_n$ is the increment of $z_n$ during the time interval ${\rm d} t$ and  $\dW_n$ are Wiener increments fulfilling $\aver{\dW_n\dW_m}=\delta_{nm}\dt$  and $\aver{\dW_n}=0$.
Note that the first term of \eref{eq:dz} the oscillator frequency $\omega_n$ is the average frequency and does not depend on time. 
The effect of the stochastic fluctuations are contained in the factors $\frac{\gamma_n}{2}z_n   \dt$ and $\sqrt{\gamma_n} z_n \dW_n$.

We now derive the equation of motion for $\tilde{\sigma}$.
To this end we consider the differential of \eref{eq:tilde(sigma)} 
\begin{eqnarray}
 \dd\tilde{\sigma}_{nm} &=& \dd z_n z^*_m+  z_n \dd z^*_m+ \dd z_n \dd z^*_m\\
&=&\Big(-i \omega_n z_n- i\sum_{m'}  \tilde{K}_{nm'}\Real{z_{m'}} -\frac{\gamma_n}{2} z_{n}   \Big) \dt\ z^*_{m}
+\sqrt{\gamma_n} z_n z_m^* \dW_n \nonumber\\
&&+ z_{n}\Big(i \omega_mz_{m}^*+i \sum_{m'}\tilde{K}_{mm'}\Real{z_{m'}} - \frac{\gamma_m}{2}z_{m}^*   \Big)  \dt 
+\sqrt{\gamma_m} z_n z_m^* \dW_m \\
&&+\sqrt{\gamma_n\gamma_m }z_n z_m^* \dW_n\dW_m\nonumber.
\end{eqnarray}
Here we have taken terms up to the first order in $\dt$ into account. 
Since $\dW$ scales like $\sqrt{\dt}$ in the first line we have included the $\dd z_n \dd z^*_m$.

We are interested in quantities that are averaged over the noise, where, as before,  the averaging is denoted by $\aver{\cdots}$. 
Defining the {\it classical density matrix}
\begin{equation}
\sigma_{nm}=\aver{\tilde{\sigma}_{nm}}
\end{equation}
we find, using \eref{eq:qm_statistics_mean}
\begin{eqnarray}
\label{eq:dd_sigma}
\dd {\sigma}_{nm} 
=&\Big(-i (\omega_n-\omega_m) - (\frac{\gamma_n}{2}+\frac{\gamma_m}{2})  \Big) \sigma_{nm} \dt\nonumber\\
&-i\sum_{m'}\frac{2V_{nm'}}{\hbar}\aver{\Real{z_{m'}}\,z_m^* }\dt\, +\, i \sum_{m'}\frac{2V_{mm'}}{\hbar}\aver{  z_n \Real{z_{m'}}} \dt\\
&+\sqrt{\gamma_n\gamma_m }\sigma_{nm}\delta_{nm} dt \nonumber
\end{eqnarray}
The term in the last line results from the averaging of the expression containing $\dW_n\dW_m$. 

If one compares \eref{eq:dd_sigma} with \eref{eq:dot_rho_nm_detail} one sees that the the term $-i (\omega_n-\omega_m)\sigma_{nm}$ corresponds to  $-(i/\hbar) (\epsilon_n-\epsilon_m)\rho_{nm}$.
The terms containing $\gamma$'s can be combined to give $\mathcal{L}[\sigma]$ where $\mathcal{L}$ is the same functional as in the quantum case.
The remaining terms contain the real part of the complex amplitude and therefore cannot be written as $\sigma_{nm}$. 
We will now first bring \eref{eq:dd_sigma}  into a form which is closer to the quantum equation and then show that in RCA they become identical.
Using  $2\,\Real{z_{m'}}=z_{m'}+z_{m'}^{*}$ we can re-write \eref{eq:dd_sigma} to obtain
\begin{eqnarray}
\label{eq:dot_sigma_final}
\dot {\sigma}_{nm} 
=&\mathcal{H}[\sigma]_{nm}+\mathcal{L}[\sigma]_{nm}
+i\sum_{\ell}\Big(\frac{V_{m\ell}}{\hbar}\aver{z_{\ell}z_n}-\frac{V_{n\ell}}{\hbar}\aver{z_{\ell}^{*}z_{m}^{*}}\Big)
\end{eqnarray}
This equation has to be compared to the quantum mechanical master equation \eref{eq:dot_rho_nm}.
Note that \eref{eq:dot_sigma_final} as it stands is not a closed system of equations for $\sigma$, due to the appearance of $\aver{z_{\ell}z_n}$ and  $\aver{z_{\ell}z_m}^{*}$.
In \ref{sec:implementation} it is  shown how we solve this equation.

We still have the freedom to normalize the classical density operator.
This will be done by the factor
\begin{equation}
\mathcal{N}=\sum_n \sigma_{nn}
\end{equation}
so that we can identify
\begin{equation}
\label{eq:identification}
\rho(t)\leftrightarrow\sigma(t)/\mathcal{N}
\end{equation}
With this we have related the classical master equation to the quantum master equation.
In the following subsection we will briefly discuss the initial state. 

\subsubsection{The initial state}
\paragraph{Pure states.}
Consider first a quantum mechanical initial state of the form
\begin{equation}
\rho^{\rm ini}=\ket{\psi^{\rm ini}}\bra{\psi^{\rm ini}}
\end{equation}
Writing the initial wave-function as $\ket{\psi^{\rm ini}}=\sum_n c_n^{\rm
  ini}\ket{\pi_n}$ with $\sum_n |c_n^{\rm ini}|^2=1$ we get for the matrix elements of the  initial density operator
\begin{equation}
\rho^{\rm ini}_{nm}= c_n^{\rm ini}(c_m^{\rm ini})^*
\end{equation}
The corresponding classical initial state is constructed by choosing 
\begin{equation}
\label{eq:init_class}
z_n^{\rm ini}=\alpha\, c_n^{\rm ini}
\end{equation}
where $\alpha$ is an overall constant which will drop out in the end, when calculating populations, coherences, etc.
Thus we have for the elements of the initial classical density matrix
\begin{equation}
\sigma^{\rm ini}_{nm}=\alpha^2 \, c_n^{\rm ini} (c_m^{\rm ini})^*
\end{equation}

\paragraph{Mixed states.}
To treat mixed states we first note that an arbitrary density matrix can be written as a weighted sum of pure states
\begin{equation}
\rho^{\rm ini}=\sum_{\beta} w_{\beta} \ket{\psi_{\beta}}\bra{\psi_{\beta}}
\end{equation}
with $\rho\ket{\psi_{\beta}}=w_{\beta}\ket{\psi_{\beta}}$.
This suggests to construct the corresponding initial classical state as
\begin{equation}
\label{equ:sigma^ini}
\sigma^{\rm ini}= \sum_{\beta} w_{\beta} \sigma_{\beta}
\end{equation}
where $(\sigma_{\beta})_{nm}=\alpha^2  (\rho_{\beta})_{nm} $.

\subsection{The realistic coupling approximation}
To investigate under which conditions the RCA will be valid we use similar arguments to those in our previous work \cite{BrEi11_051911_,BrEi11__a}.
To this end we will consider not the density matrix equations but the equivalent stochastic equations.
Expanding the quantum wave function as
\begin{equation}
\ket{\Psi(t)}=\sum_{n}c_n(t) \ket{\pi_n}
\end{equation}
in the quantum case one has from \eref{eq:Ham_deph_stoch}.
\begin{equation}
\dot{c}_n=-i (\epsilon_n+h_n(t))c_n -i \sum_{m}V_{nm}c_{m}
\end{equation}
which has to be compared to the classical equation  \eref{eq:dot_z_final} which can be written as
\begin{equation}
\label{eq:dot_z_final_appRCA}
\dot{z}_n=-i (\omega_n+w_n(t)) z_n - i\sum_m \frac{2 V_{nm}}{\hbar} \Real (z_m)
\end{equation}
where the stochastic processes can be chosen to be $w_n(t)=h_n(t)/\hbar$ and as before $\omega_n=\epsilon_n/\hbar$.

In Ref.~\cite{BrEi11_051911_}, where $h(t)\equiv 0$, it was noted that under the conditions that 
\begin{eqnarray}
\label{eq:RCA_V<eps}
|V_{nm}|/\hbar &\ll& \omega_n\\
\label{eq:RCA_eps_n-eps_m<eps}
\big|\omega_n-\omega_m\big| &\ll& \omega_n
\end{eqnarray}
 the classical equations accurately describe the results obtained from the Schr\"odinger equation.
Then also the corresponding classical master equation should give results which agree with those obtained from  the quantum master equation.

In the present case, where $h(t)\ne 0$,  we expect that the difference between the quantum and the classical evolution is small if the same conditions for the time dependent frequencies are fullfilled, namely that
\begin{eqnarray}
\label{eq:RCA_timedep_abs}
|V_{nm}|/\hbar& \ll \omega_n+w_n(t)\\
\omega_n &\gg \big|\omega_n+w_n(t)-(\omega_m+w_m(t))\big|
\label{eq:RCA_timedep_diff}
\end{eqnarray}
for most times $t$.
Since $w_n(t)$ can take negative values we see from Eq.~(\ref{eq:RCA_timedep_abs}) that the fluctuations have to be small compared to to the average frequency $\omega_n$.
The second equation (\ref{eq:RCA_timedep_diff}) states that the frequency difference between different sites has to be small compared to the mean frequency.
 
For the situation considered in the present work we use $\gamma_n$ as a rough measure of the magnitude of the frequency variations. 
Then both equations (\ref{eq:RCA_timedep_abs}) and (\ref{eq:RCA_timedep_diff}) lead to the estimate
\begin{eqnarray}
\label{eq:RCA_timedep_abs_gamma}
\gamma_n &\ll& \omega_n
\end{eqnarray}
If this condition and (\ref{eq:RCA_V<eps}) and (\ref{eq:RCA_eps_n-eps_m<eps}) hold, then we expect the RCA to be a good approximation.
Note, that these assumptions have to be fullfilled also for the quantum mechanical model employed to be meaningful. That is the energy changes experienced by a given molecule due to coupling to other molecules and the environment must be small compared to the unperturbed molecule transition energy.

In \sref{sec:RCA_test} we will investigate the range of validity of this approximation in more detail.
In particular we will show that for typical parameters used in the photosynthetic FMO complex the classical equations give a good description to the quantum mechanical exciton dynamics.

\section{Comparison of quantum and classical results}
\label{sec:RCA_test}

Here we compare the results obtained using the classical ``master equation'' \eref{eq:dot_sigma_final} with those obtained from the quantum one \eref{eq:dot_rho_nm}. 
The classical equation \eref{eq:dot_sigma_final} is solved using the method
described in  \ref{sec:implementation}.

First we will consider the case of a linear chain and afterwards discuss the FMO complex.
For both cases we investigate the populations as well as the coherences between different sites.

\subsection{The linear chain}
As a first example we consider the  standard case  of a linear chain where all transition energies $\epsilon_n$ and dephasing rates $\gamma_n$ are identical. For simplicity we take only nearest neighbor interaction into account.
This interaction, denoted by $V$ is taken as the unit of energy.

\begin{figure}[tp]
\center
\includegraphics[width=14cm]{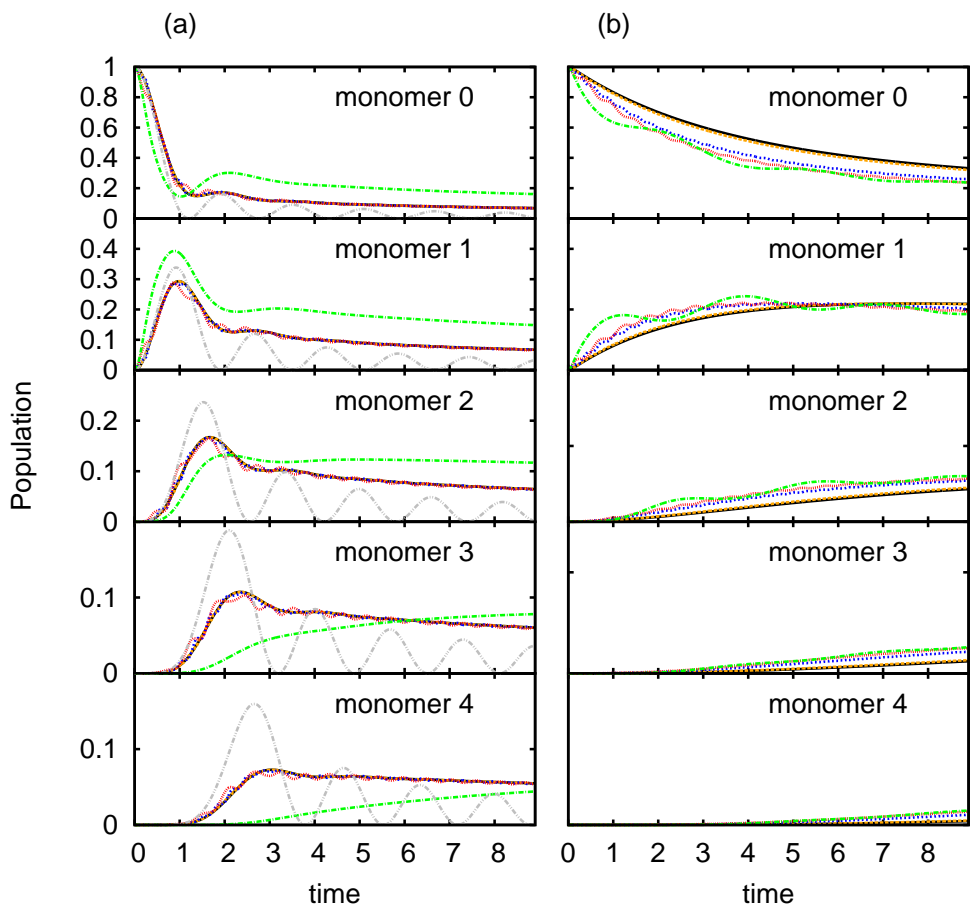}
\caption{\label{fig:lin_chain_pops}
Time-dependent populations of a linear chain when the excitation is localized initially on one monomer ``0''. The transport is symmetric w.r.t.\ this monomer. 
Left: $\gamma= V$ (for comparison also the analytic  $\gamma = 0$ result is shown as thin, gray curve). Right: $\gamma=20\, V$.
Bold black: exact quantum  calculation. 
The colored curves are results from the classical calculations for different $\epsilon$.
Orange: $\epsilon=40\; V$. 
Blue: $\epsilon=10\; V$. 
Red: $\epsilon=6\; V$.
Green: $\epsilon=1\; V$.
Time is in units of $V/\hbar$.
}
\end{figure}

In the results shown in  \fref{fig:lin_chain_pops} we have used as initial condition a state where the excitation is initially localized on a single monomer which we denote by '0' \footnote{For the calculation we used a chain of 29 sites and started at site 14.}.
In each column a fixed $\gamma $ is chosen and the transition energy $\epsilon$ is varied. 
For reference purposes, in the left column, the quantum solution for $\gamma=0$ is also shown. Here the coherence is maximal and the probability of EET is given by the square of a Bessel function \cite{Me58_647_} which is oscillatory in time, reaching zero at the zeroes of the Bessel function.
As $\gamma$ is increased one sees that the oscillations are damped for $\gamma = V$ (left column of Fig.~\ref{fig:lin_chain_pops}). 
When $\gamma = 20V$ (right column) the oscillations are washed out completely and the populations change monotonically in the quantum case.
The exact quantum results are shown as solid black lines.

To investigate the validity of the RCA we show classical solutions for various values of the transition energy $\epsilon= \hbar\omega$.
In particular we have chosen $\epsilon= 40,30,10,6$, and $1$ in units of $V$.
For the values  $\epsilon= 40,30,10$ Eq.~(\ref{eq:RCA_V<eps}) is fulfilled while it definitely does not hold for $\epsilon=1$. 
Note that for our choice of identical transition energies Eq.~(\ref{eq:RCA_eps_n-eps_m<eps}) is trivially fulfilled.
For the case $\gamma=1$ we have $\epsilon/\gamma=\epsilon/V$ and the inequality 
(\ref{eq:RCA_timedep_abs_gamma}) is fulfilled whenever (\ref{eq:RCA_V<eps}) is fulfilled.
However for $\gamma =20$ the inequality (\ref{eq:RCA_timedep_abs_gamma}) is not fulfilled for $\epsilon=10,6,1$. 
For this case one expects to see the influence of the fluctuating transition energies on the RCA.
 
These expectations are met by the numerical results shown in Fig.~\ref{fig:lin_chain_pops}.
We see that for the case $\gamma=1$ for $\epsilon \ge 10$ the quantum and classical results are nearly indistinguishable. 
Even for  a ratio $\epsilon/\gamma =6$ (red curve) there is still quite good agreement.
For even smaller $\epsilon/\gamma$ the deviations become more pronounced as exemplified by  $\epsilon/\gamma=1$ (green curve). 
For the large value $\gamma=20$ the classical results are indistinguishable from the quantum ones for the cases 
$\epsilon\ge 30$ (where both inequalities (\ref{eq:RCA_timedep_abs_gamma}) and (\ref{eq:RCA_V<eps}) hold).  
For all other values (where $\epsilon<\gamma$) there are clear deviations from the quantum result.

\begin{figure}[tp]
\psfrag{rho1,2}{$|\rho_{0,1}|$}
\psfrag{rho1,3}{$|\rho_{0,2}|$}
\psfrag{rho1,4}{$|\rho_{0,3}|$}
\includegraphics[width=17cm]{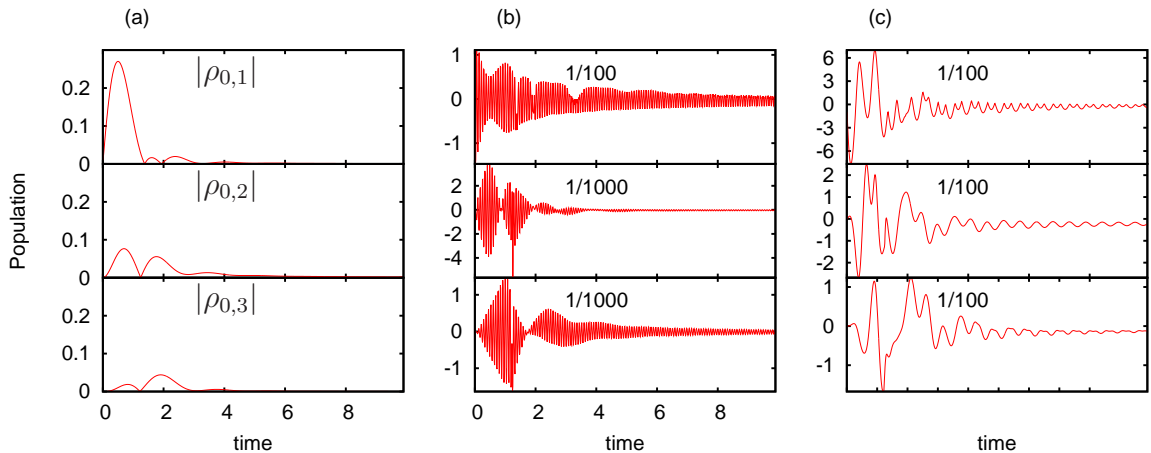}
\caption{\label{fig:lin_chain_cor}
Left column: Quantum coherences  $|\rho_{0,1}|$, $|\rho_{0,2}|$ and $|\rho_{0,3}|$  of a linear chain when initially the excitation is localized on one monomer ``0''.
The dephasing rate is $\gamma=V$.
Middle column: Differences between  classical and quantum results for the case $\epsilon/\gamma=40$.
Right column: Differences between  classical and quantum results for the case $\epsilon/\gamma=6$.
Note the different scalings in columns (b) and (c).
Time is in units of $V/\hbar$.
}
\end{figure}

Not only the populations but also the inter-site coherences obtained from the classical 'density matrix' are in good agreement with the quantum mechanical coherences. This is demonstrated in \fref{fig:lin_chain_cor} for the case $\gamma=V$ where  in the left column  for a linear chain the time-dependent absolute values of the coherences between site '0' and site '1' (upper row), site '2' (middle row) and site '3' (lower row), obtained from full quantum calculations, are shown. Oscillations in the coherences are evident.
Since deviations between classical and quantum results are not always easy to distinguish, in the middle column the \emph{differences} between the absolute values of the quantum result and the classical calculation for the case $\epsilon/\gamma=40$ are shown.
One sees fast fluctuations in the differences  but they are always at least  two orders of magnitude smaller than the magnitude of the exact coherences.
Upon decreasing the ratio $\epsilon/\gamma$ the fluctuations in the difference become larger and slower but never exceed more than a few percent.
An example is shown in the right column for $\epsilon/\gamma=6$.

\subsection{The photosynthetic FMO complex}
The  simple HRS model can be  used to gain insight into the dynamics of excitation energy transfer in photysynthetic systems, although for a more realistic description a more detailed treatment of the environment  would be necessary \cite{WePuPr00_5825_,ARXIV_Sangwoo,RiRoSt11__,OlCoLi11__}.

Here we show, for the case of pure dephasing (with $\gamma_n=\gamma$), that for values of the transition energies and interactions between the BChl molecules that are typical for such systems the classical equation gives results in very good agreement with those of the quantum one.

In the following we use the energies and couplings for {\it c.\ tepidum} as given in Ref.~\cite{AdRe06_2778_} (the site energies are taken from the trimeric structure of table 4 and the intersite couplings are taken from the fourth column of table 1 of that paper).
We note that for our comparison the exact values are not important, therefore we did not consider more recent values \cite{ScMuMa11_93_,ARXIV_Sangwoo,OlCoLi11__}.
One would get similar results as presented below for slightly different parameters and also when treating the full trimeric FMO system with 24 BChl molecules.

\begin{figure}
\includegraphics[width=15 cm]{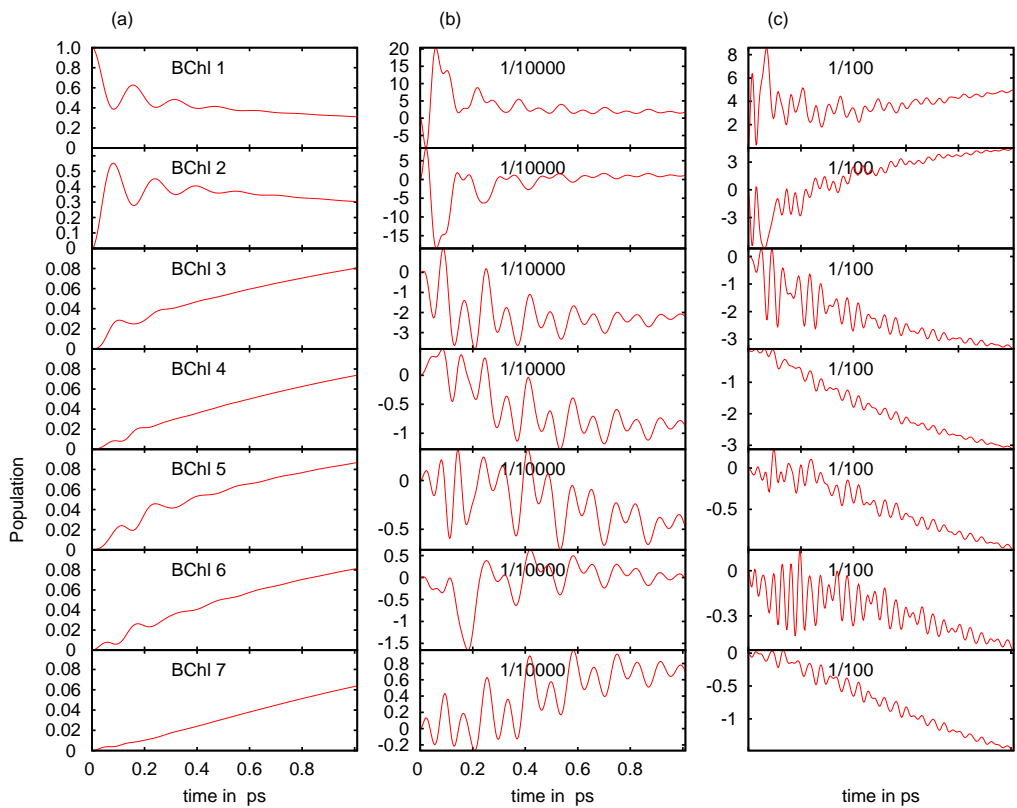}
\caption{\label{fig:FMO_Pops}a) Populations of the BChls as function of time when excitation is initially localised on BChl 1 obtained from the full quantum mechanical calculations. b) Differences between the quantum result and the classical calculation for the actual transition energies. c) The transition energies are reduced by  12000 cm$^{-1}$ .
Note the different scales of the y axis and the scaling factors which are indicated in the figures.
}
\end{figure}

In  \fref{fig:FMO_Pops} the time dependence of the excitation probabilities of the Bchl molecules  is shown. The left column shows the results from the full quantum calculation.
The middle and the right columns show the differences between the quantum result and the solution of the classical equation. In the middle column the transition energies are taken as given in Ref.~\cite{AdRe06_2778_}. They are in the order of 12000 cm$^{-1}$ which is much larger than the coupling between the BChls (which is in the order of 100 cm$^{-1}$) and also much larger than the energy differences between the transition energies (which are also a few hundred wavenumbers).  
As expected the the deviations from the exact quantum result are quite small (on the order of $0.1\%$).
Even if one reduces the transition energies by 12000 cm$^{-1}$ (which results in completely unrealistic transition energies of the order of a few hundred wavenumbers) the deviations are still only on the order of $10\%$. 

\begin{figure}
\psfrag{rho31}{$\rho_{31}$}
\psfrag{rho32}{$\rho_{32}$}
\psfrag{rho33}{$\rho_{33}$}
\psfrag{rho34}{$\rho_{34}$}
\psfrag{rho35}{$\rho_{35}$}
\psfrag{rho36}{$\rho_{36}$}
\psfrag{rho37}{$\rho_{37}$}
\includegraphics[width=15 cm]{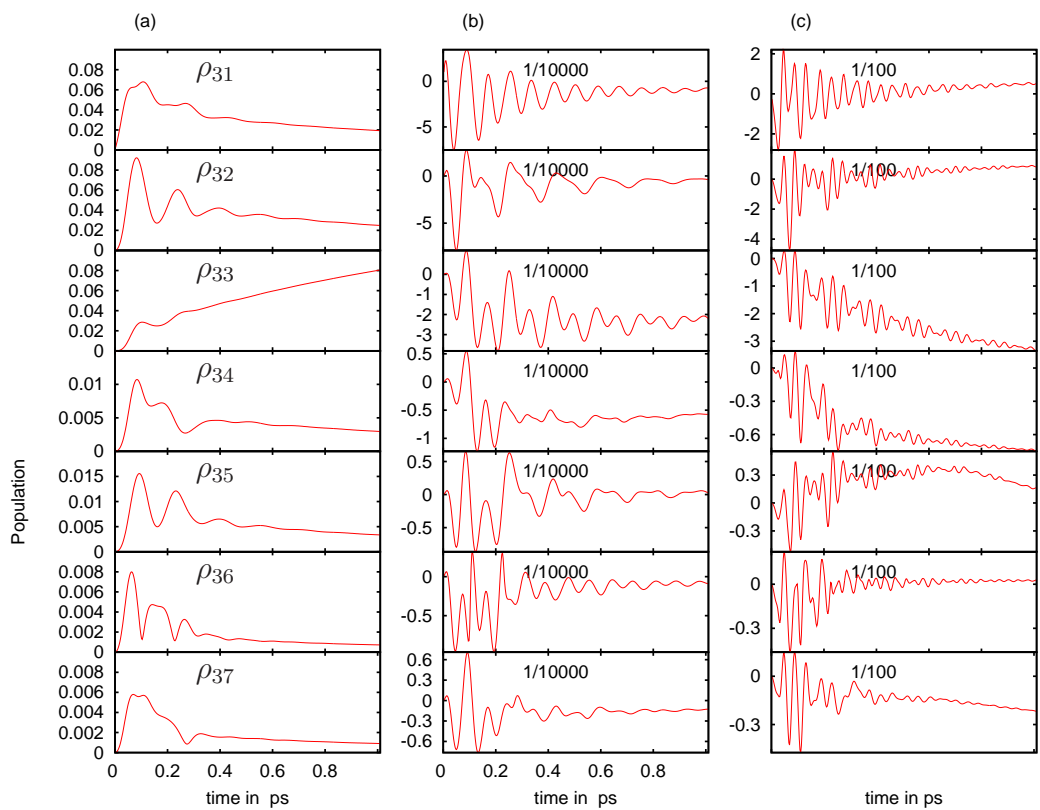}
\caption{\label{fig:FMO_Cor}As \fref{fig:FMO_Pops} but now for the coherences of BChl 3 with all the other BChls.
Note that in the third row also the diagonal term i.e.\ the population of BChl 3 is shown. Note the scaling factors in column (b) and (c).
}
\end{figure}

The agreement between the classical coherences and the quantum coherences is of a similar order.
This is exemplified by the results shown in \fref{fig:FMO_Cor}.
The absolute differences are of the same order of magnitude as the differences in the populations. However, since the coherences are smaller, the relative error is somewhat larger.


\section{Conclusions}
In the present paper we have extended our previous investigation on the correspondence between quantum mechanical and classical EET to include coupling to an environment.
In particular we have demonstrated that it is possible to derive a master equation for the classical amplitudes which in the RCA reproduces the corresponding quantum master equation.
This has been shown explicitly for the case of an environment that leads to pure dephasing.
As shown by HRS the corresponding quantum master equation can be obtained from an average over a stochastic Schr\"odinger equation with real Markovian noise that alters the transition energies of the monomers.   
We used this stochastic representation to relate the quantum Schr\"odinger equation to a classical equation for coupled harmonic oscillators with frequencies that have the same stochastic properties as the quantum transition energies.
We then found that the classical results reproduce the quantum results when the fluctuations in the transition energies are small compared to the transition energy.
This has been demonstrated explicitly by considering both a linear chain and the FMO complex.

Although we have made this demonstration for real Gaussian Markovian noise it is clear that the same argument will also hold for more general stochastic processes for the transition energies, which may be non-Gaussian or non-Markovian (of course also deterministic functions and fluctuations of the couplings between different sites can be treated in this manner).
 
In this respect it is worth mentioning that such fluctuating  site transition energies arise in molecular dynamics/quantum chemistry simulations performed on photosynthetic complexes \cite{DaKoKl02_031919_,ARXIV_Sangwoo,OlJaLi11_8609_}.
Thus the transport and coherence properties obtained from  such studies could be reproduced by a purely classical model. 

 The  results presented demonstrate that classical oscillators can be used to simulate the transport and coherence properties of molecular aggregates and in particular those of photosynthetic systems. In a recent article \cite{BrEi11__a} we have discussed how classical electrical LC circuit oscillators can be used to  mimic coupled quantum two level systems. 
From the foregoing it is clear that by modulating the frequencies of the classical oscillators in an appropriate way the classical LC oscillators can be used to simulate the quantum aggregate. 
Such classical simulations could be used to compare with recently proposed simulations using superconducting qubits \cite{ARXIV_Mostame}.

\ack
Thanks are due to Jan Roden for providing many useful fortran routinesand to Gerhard Ritschel and Sebstian M\"obius for careful reading of the manuscript.
AE thanks Al\'{a}n Aspuru-Guzik for hospitality.
Financial support by the DFG under Contract No. Ei~872/1-1 is acknowledged. 

\appendix

\section{Solution of the classical equation}
\label{sec:implementation}

In this section we describe how we solve the classical equation \eref{eq:dot_sigma_final}.
To this end we introduce the auxiliary matrices
\begin{eqnarray}
\rA_{nm}&=\aver{\tilde{x}_n\,\tilde{x}_m}
\label{eq:rA_ini}\\
\rB_{nm}&=\aver{\tilde{p}_n\,\tilde{p}_m}
\label{eq:rB_ini}\\
\rC_{nm}&=\aver{\tilde{x}_n\,\tilde{p}_m}
\label{eq:rC_ini}
\end{eqnarray}
With this the classical density operator can be written as
\begin{equation}
\sigma_{nm}=\rA_{nm}+\rB_{nm}+i (\rC_{mn}-\rC_{nm})
\end{equation}

From the evolution equations of $\tilde{x}_n$  and  $\tilde{p}_n$ we can then derive a set of coupled equations:
\begin{eqnarray}
\label{eq:dot_components_sigma}
\dot{\rA}_{nm}=& \omega_n \rC_{mn}+\omega_m \rC_{nm}+\mathcal{L}[\rA]_{nm}\\
\dot{\rB}_{nm}=&-( \omega_n \rC_{nm}+\omega_m \rC_{mn})+\mathcal{L}[\rB]_{nm}
-\sum_{\ell}(\frac{2V_{n\ell}}{\hbar}\rC_{\ell m}+\frac{2V_{m\ell}}{\hbar} \rC_{\ell n})\\
\label{eq:dot_components_sigma_3}
\dot{\rC}_{nm}=& \omega_n \rB_{nm}-\omega_m \rA_{nm}+\mathcal{L}[\rC]_{nm}
-\sum_{\ell}\frac{2V_{m\ell}}{\hbar} \rA_{n \ell }
\end{eqnarray}

\subsection{The initial state}

In the following we set for convenience the normalisation factor $\alpha$, introduced  in \eref{eq:init_class}, to $\alpha=1$.
 For each $\beta$ in \eref{equ:sigma^ini} we make the identification $z^{\beta}_n=c^{\beta}_n$.
Then we construct the components $\rA^{\beta}$, $\rB^{\beta}$ and $\rC^{\beta}$ from the equations \eref{eq:rA_ini}-\ref{eq:rC_ini}. 
This gives
\begin{eqnarray}
\sigma^{ini}_{nm}=&\sum_{\beta} w_{\beta} \sigma^{\beta}_{nm}\\
=&\rA^{\rm ini}_{nm}+\rB^{\rm ini}_{nm}+i (\rC^{\rm ini}_{mn}-\rC^{\rm ini}_{nm})
\end{eqnarray} 
with $\rA^{\rm ini}_{nm}=\sum_{\beta} w_{\beta}\rA^{\beta}_{nm}$,
 $\rB^{\rm ini}_{nm}=\sum_{\beta} w_{\beta}\rB^{\beta}_{nm}$ and 
$\rC^{\rm ini}_{nm}=\sum_{\beta} w_{\beta}\rC^{\beta}_{nm}$.
Thus the initial vector $(\rA^{\rm ini},\rB^{\rm ini},\rC^{\rm ini})$, needed for the propagation with \eref{eq:dot_sigma_final}, can be written as a sum of the vectors $(\rA^{\beta},\rB^{\beta},\rC^{\beta})$.
The linearity of \eref{eq:dot_components_sigma}-(\ref{eq:dot_components_sigma_3}) then guarantees that $\sigma(t)$ can be obtained at later times.

\subsection{Comparison with the qm-equation}
Similarly as for the classical equation one can also re-write the quantum equation, giving further insight into the RCA approximation.
In the quantum case we define
\begin{eqnarray}
\rAqm_{nm}&=\Real{c}_n\Real{c}_m\\
\rBqm_{nm}&=\Imag{c}_n\Imag{c}_m\\
\rCqm_{nm}&=\Real{c}_n\Imag{c}_m
\end{eqnarray}
With this the density operator can be written as
\begin{equation}
\rho_{nm}=\rAqm_{nm}+\rBqm_{nm}+i (\rCqm_{nm}+\rCqm_{mn})
\end{equation}

We can then derive the set of coupled equations:
\begin{eqnarray}
\fl {\dotrAqm}_{nm}= \omega_n \rCqm_{mn}+\omega_m \rCqm_{nm}+\mathcal{L}[\rAqm]_{nm}
+\sum_{\ell}(\frac{V_{n\ell}}{\hbar}\rCqm_{m \ell }+\frac{V_{m\ell}}{\hbar} \rCqm_{n \ell })
\\
\fl{\dotrBqm}_{nm}=-(\omega_n \rCqm_{nm}+\omega_m \rCqm_{mn})+\mathcal{L}[\rBqm]_{nm}
-\sum_{\ell}(\frac{V_{n\ell}}{\hbar}\rCqm_{\ell m}+\frac{V_{m\ell}}{\hbar} \rCqm_{\ell n})
\\
\fl {\dotrCqm}_{nm}= \omega_n \rBqm_{nm}-\omega_m \rAqm_{nm}+\mathcal{L}[\rCqm]_{nm}
+\sum_{\ell}(-\frac{V_{m\ell}}{\hbar} \rAqm_{n \ell}+\frac{V_{n\ell}}{\hbar} \rBqm_{ \ell m})
\end{eqnarray}
One sees that compared to the classical case the terms containing the interaction between the monomers are more symmetrically distributed among the individual equations.

%
%


\vspace{1cm}
\section*{References}

\begin{thebibliography}{10}
\providecommand{\url}[1]{\texttt{#1}}
\providecommand{\urlprefix}{URL }
\expandafter\ifx\csname urlstyle\endcsname\relax
  \providecommand{\doi}[1]{doi:\discretionary{}{}{}#1}\else
  \providecommand{\doi}{doi:\discretionary{}{}{}\begingroup
  \urlstyle{rm}\Url}\fi
\providecommand{\eprint}[2][]{\url{#2}}

\bibitem{BrEi11_051911_}
J.~S. Briggs and A.~Eisfeld; Phys. Rev. E \textbf{83} 051911 (2011).

\bibitem{SaIsFl10_462_}
M.~Sarovar, A.~Ishizaki, G.~R. Fleming and K.~B. Whaley; Nat Phys \textbf{6}
  462 (2010).

\bibitem{HaRe71_253_}
H.~Haken and P.~Reineker; Z. Phys. \textbf{249} 253 (1971).

\bibitem{HaSt73_135_}
H.~Haken and G.~Strobl; Z. Phys. \textbf{262} 135 (1973).

\bibitem{LiVaGr97_7343_}
V.~Liuolia, L.~Valkunas and R.~van Grondelle; J. Phys. Chem. B \textbf{101}
  7343 (1997).

\bibitem{RaKnKe79_197_}
T.~S. Rahman, R.~S. Knox and V.~M. Kenkre; Chem. Phys. \textbf{44} 197 (1979).

\bibitem{ReWaNe93_715_}
P.~Reineker, C.~Warns, T.~Neidlinger and I.~Barvík; Chemical Physics
  \textbf{177} 715 (1993).

\bibitem{SuEnRe06_337_}
C.~Supritz, A.~Engelmann and P.~Reineker; Journal Of Luminescence \textbf{119}
  337 (2006).

\bibitem{CaChDa09_105106_}
F.~Caruso, A.~W. Chin, A.~Datta, S.~F. Huelga and M.~B. Plenio; J. Chem. Phys.
  \textbf{131} 105106 (2009).

\bibitem{ReMoKa09_033003_}
P.~Rebentrost, M.~Mohseni, I.~Kassal, S.~Lloyd and A.~Aspuru-Guzik; New Journal
  of Physics \textbf{11} 033003 (12pp) (2009).

\bibitem{Ca85_101_}
V.~\v{C}\'{a}pek; Z. Phys. B \textbf{60} 101 (1985).

\bibitem{BlSi78_3589_}
A.~Blumen and R.~Silbey; J. Chem. Phys. \textbf{69} 3589 (1978).

\bibitem{SzBa86_179_}
V.~Szöcs and I.~Barvík; Journal of Theoretical Biology \textbf{122} 179
  (1986).

\bibitem{WuKn98_359_}
M.~Wubs and J.~Knoester; J. Lumin. \textbf{76-7} 359 (1998).

\bibitem{BrEi11__a}
J.~Briggs and A.~Eisfeld; arXiv:1104.4158v1 [quant-ph]  (2011).

\bibitem{Ku54_935_}
R.~Kubo; J. Phys. Soc. Jpn. \textbf{9} 935 (1954).

\bibitem{Fo78_179_}
R.~F. Fox; PHYSICS REPORTS \textbf{48} 179 (1978).

\bibitem{Ka81_175_}
N.~G. van Kampen; Journal of Statistical Physics \textbf{24} 175 (1981).

\bibitem{HaEz80_41_}
H.~Hasegawa and H.~Ezawa; Prog. Theor. Phys. Suppl. \textbf{69} 41 (1980).

\bibitem{Me58_647_}
R.~E. Merrifield; J. Chem. Phys. \textbf{28} 647 (1958).

\bibitem{WePuPr00_5825_}
M.~Wendling, T.~Pullerits, M.~A. Przyjalgowski, S.~I.~E. Vulto, T.~J. Aartsma,
  R.~van Grondelle and H.~van Amerongen; The Journal of Physical Chemistry B
  \textbf{104} 5825 (2000).

\bibitem{ARXIV_Sangwoo}
S.~Shim, P.~Rebentrost, S.~Valleau and A.~Aspuru-Guzik; arXiv:1104.2943v1
  [quant-ph] .

\bibitem{RiRoSt11__}
G.~Ritschel, J.~Roden, W.~T. Strunz and A.~Eisfeld; arXiv:1106.5259v1
  [quant-ph]  (2011).

\bibitem{OlCoLi11__}
C.~Olbrich, T.~la~Cour~Jansen, J.~Liebers, M.~Aghtar, J.~Str\"umpfer,
  K.~Schulten, J.~Knoester and U.~Kleinekath\"ofer; J. Phys. Chem. B, Accepted,
  DOI: 10.1021/jp202619a .

\bibitem{AdRe06_2778_}
J.~Adolphs and T.~Renger; Biophys J \textbf{91} 2778 (2006).

\bibitem{ScMuMa11_93_}
M.~Schmidt am~Busch, F.~M\"uh, M.~E.-A. Mohamed and T.~Renger; Journal of
  Physical Chemistry Letters \textbf{{2}} 93 (2011).

\bibitem{DaKoKl02_031919_}
A.~Damjanovi\'{c}, I.~Kosztin, U.~Kleinekath\"ofer and K.~Schulten; Phys. Rev.
  E \textbf{65} 031919 (2002).

\bibitem{OlJaLi11_8609_}
C.~Olbrich, T.~L.~C. Jansen, J.~Liebers, M.~Aghtar, J.~Str\"umpfer,
  K.~Schulten, J.~Knoester and U.~Kleinekath\"ofer; The Journal of Physical
  Chemistry B \textbf{115} 8609 (2011).

\bibitem{ARXIV_Mostame}
S.~Mostame, P.~Rebentrost, D.~I.~Tsomokos and A.~Aspuru-Guzik; arXiv:1106.1683v1
  [quant-ph] .

\end{thebibliography}
\end{document}